\documentclass[aps,pra,twocolumn,groupedaddress,floatfix]{revtex4-1}

\usepackage{graphicx}
\usepackage{dcolumn}
\usepackage{bm}
\usepackage{amsmath}
\usepackage{xcolor}
\usepackage{braket}

\usepackage{bm}
\usepackage[utf8]{inputenc}
\usepackage[T1]{fontenc}
\usepackage{hyperref}
\usepackage[all]{hypcap}
\hypersetup{
breaklinks=true,
    colorlinks=true,
    linkcolor=blue,
    citecolor=blue,
    filecolor=magenta,
    urlcolor=cyan,
}

\usepackage{amsfonts}
\usepackage{blkarray}
\usepackage{amssymb}
\usepackage{multirow}
\usepackage{mathrsfs}

\newcommand{\cp}{{\cal P}}
\newcommand{\cz}{{\cal Z}}
\newcommand{\vdn}{ \Phi_V }
\newcommand{\sypa}{ \Phi_{\rm sym} }

\newcommand{\cg}{{\cal G}}

\begin{document}

\title{Exact closed-form analytic wave functions in two dimensions: Contact-interacting
fermionic spinful ultracold atoms in a rapidly rotating trap} 

\author{Constantine Yannouleas}
\email{Constantine.Yannouleas@physics.gatech.edu}
\author{Uzi Landman}
\email{Uzi.Landman@physics.gatech.edu}

\affiliation{School of Physics, Georgia Institute of Technology,
             Atlanta, Georgia 30332-0430}

\date{12 January 2021, Letter, Phys. Rev. Research, {\bf in press}}

\begin{abstract}
Exact two-dimensional analytic wave functions for an arbitrary number $N$ of contact-interacting
lowest-Landau-level (LLL) spinful fermions are derived with the use of 
combined numerical and symbolic
computational approaches via analysis of exact Hamiltonian numerical diagonalization data.
Closed-form analytic expressions are presented for two families of zero-interaction-energy
states at given total angular momentum and total spin $0 \leq S \leq N/2$ in the neighborhood of
the $\nu=1$ filling, covering the range from the maximum density droplet to the first quasihole.
Our theoretical predictions for higher-order spatial and  momentum correlations reveal intrinsic 
polygonal, multi-ring crystalline-type structures, which can be tested with ultracold-atom 
experiments in rapidly rotating traps, simulating quantum Hall physics (including quantum LLL 
skyrmions).
\end{abstract}

\maketitle

\section{Introduction}
\label{intr}

Exact analytic solutions for the quantum many-body problem, 
whether in a closed-form algebraic expression or in the form of the Bethe ansatz,
are highly coveted and sought-after; however, they are available only for a few cases.
Among this select group (for early pioneering studies see 
Refs.~\cite{gira60,lieb63,mcgu64,calo71,suth71,mattisbook,suthbook}), 
one-dimensional (1D) assemblies 
of strongly contact-interacting ultracold atoms have attracted much attention in the last few 
years \cite{gira01,gira07,gira10,guan09,cui14,deur14,zinn14,bruu15,cui16,astr20}, 
motivated by rapid experimental advances in the field of trapped ultracold atoms
that allow direct verification of theoretical results. 
In this context, {\it in-situ\/} and time-of-flight single-atom measurements of 
real-space and momentum-space higher-order correlations, respectively, hold a great promise
\cite{grei02,gerb05,gerb05.2,bloc15,ott16,hodg17,clem18,prei19,berg19,clem19,holt20,
yann07.2,yann19,yann20.2}.  

Here we derive {\it closed-form exact analytic\/} wave functions (EAWFs) for two-dimensional (2D) 
systems of spinful contact-interacting lowest-Landau-level (LLL) fermions that simulate fractional
quantum Hall (FQH) physics \cite{sanp07,coop08,popp04,hazz08,geme10,caru18,palm20,yann20} 
with trapped ultracold atoms. We first introduce a novel approach for the extraction of EAWFs 
from the digital information provided via numerical exact-diagonalization (i.e., the configuration 
interaction, CI \cite{shav98,yann20,palm20,yann07.2}) of the many-body LLL Hamiltonian.
Subsequently, we present illustrative examples, showing that such 
EAWFs exhibit intrinsic geometric structures (ultracold Wigner molecules, UCWMs) in their 
higher-order correlations, in line with earlier findings using numerical CI solutions (see, e.g.,
Ref.~\cite{yann20}). The compact EAWFs enable consideration of larger assemblies
compared to the CI-computed UCWMs \cite{yann20}.

Starting with the Laughlin trial wave function \cite{laug83}, compact algebraic forms 
have been extensively considered \cite{coop08,simochap,yoso98,halp83,jainbook,brey96,roug14,lieb20}
as approximations to the exact diagonalization solutions, both for electrons in
semiconductors \cite{simochap,yoso98,halp83,jainbook,brey96} and for ultracold bosons in 
rotating traps \cite{coop08,roug14,lieb20}. In several instances, like the Laughlin wave 
functions, it was shown that the variational trial functions \cite{simochap,jainbook,coop08} 
may be exact solutions, with zero-interaction energy (0IE states), 
of specific short-range pseudopotential-type parent Hamiltonians \cite{simochap,coop08}.

Because of the fermionic statistics, this paper relates to electronic 
2D quantum LLL skyrmions \cite{brey96,jain96,note1,abol97,girv98},
multicomponent quantum Hall systems \cite{girv98},
2D anyons \cite{caru18}, and rotating electronic \cite{yann02,yann03,yann04,yann07}
and ultracold-atom \cite{yann06,yann07.2,yann20} Wigner molecules. Experimental realization of 
such 2D systems (including bosonic analogs \cite{popp04,hazz08,geme10,caru18}) 
with a few ultracold fermionic atoms (e.g., $^6$Li) in rapidly rotating harmonic traps is 
currently pursued \cite{palm20}. Importantly, unlike the skyrmion wave functions used in the 
literature \cite{brey96,jain96,note1,abol97,yang06}, which are not eigenstates of the total spin 
(see particularly Ref.~\cite{abol97}, the Appendix,  and the Supplemental Material (SM) 
\cite{supp}), the EAWFs introduced here provide total-spin preserving symmetric polynomials for 
the quantum LLL skyrmions; for other spin-preserving polynomials (restricted to the spin-singlet 
state), see Ref.~\cite{note2}.

\section{Methodology}

Extensions of Girardeau's mapping between impenetrable bosons
and non-interacting spinless fermions \cite{gira60}, and similar mappings 
\cite{gira01,gira10} applied to spinful and spin-parallel fermions, led to the formulation of 
a hard-core boundary condition for strongly-repelling 1D fermions \cite{guan09,cui14}. This 
entails vanishing of the many-body wave functions when two fermions with antiparallel spins 
are at the same position (in addition to the vanishing for parallel spins due to the Pauli 
exclusion principle). Concomitant of this condition is the appearance of 0IE states.

\begin{figure}[t]
\centering\includegraphics[width=7.8cm]{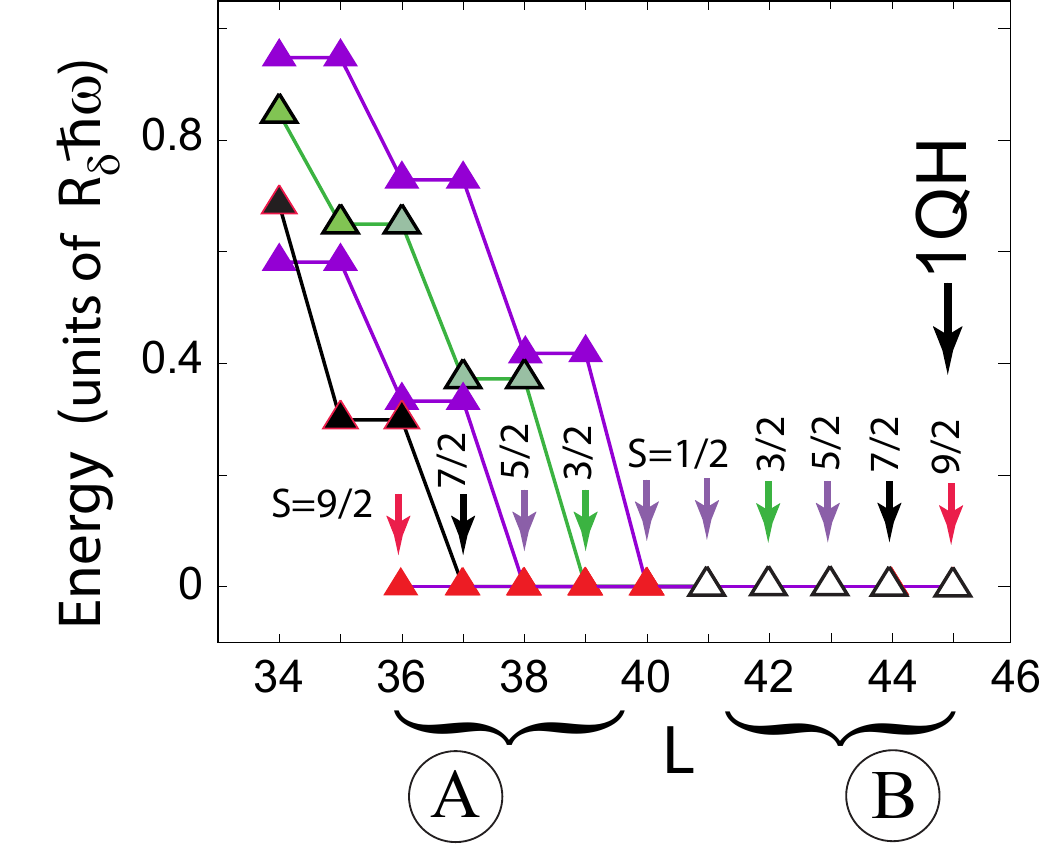}
\caption{
CI-calculated, relative-ground-state LLL energy spectra for $N=9$ fermions associated with the 
contact-interaction term only; see third term of $H_{\rm LLL}$ in Eq.~(\ref{H_lll}).
Spectra in a given spin sector $S=1/2, 3/2, 5/2, 7/2$ and $S=9/2$ are expilcitly denoted.
The spectra were calculated for $S_z=1/2$; however, they are independent of the precise value of 
$S_z$. The 0IE states of family $A$ are colored in red; those of family $B$ are colored in white. 
The first quasi-hole state is also explicitly denoted. $L$ is the total angular momentum. 
}
\label{fspec}
\end{figure}

In CI calculations and for a given number $N$ of spinful LLL fermions, 
the 0IE states emerge in each spin sector $(S, S_z)$; see Fig.~\ref{fspec} 
for the case of $N=9$ LLL fermions interacting with a $\sum_{i<j} \delta(z_i-z_j)$ two-body 
potential, where $z_i=x_i+i y_i$ (with $i=1,2, \ldots, N$). 
Interest in such 0IE states arises from: 
(i) They can be prepared experimentally \cite{palm20}. (For bosons, the experimental
expectations for 0IE states include fittingly the $\nu=1/2$ bosonic Laughlin state
\cite{popp04,hazz08,dali11}.) 
(ii) They represent microscopic states that describe quantum LLL skyrmions \cite{brey96,note1}.
(iii) The Laughlin states are 0IE eigenstates associated with short-range pseudopotential-type
Hamiltonians \cite{hald83,simochap}. (iv) For fully polarized fermions, 0IE states have been
associated with the gapless edge excitations of the Laughlin droplet \cite{wen92} in extended
semiconductor samples.

The many-body Hamiltonian describing ultracold neutral atoms in a rapidly rotating trap 
\cite{popp04,yann06,yann07.2,coop08,hazz08,palm20} is given by
\begin{equation}
\frac{ H_{\rm LLL} }{\hbar \omega}=N + (1-\frac{\Omega}{\omega}) L
+2 \pi R_\delta \sum_{i < j}^{N} \delta(z_i-z_j),
\label{H_lll}
\end{equation}
where $\omega$ and $\Omega$ are, respectively, the parabolic trapping and rotational frequencies 
of the trap, and $L$ denotes the total angular momentum, $L=\sum_{i=1}^N\ l_i$, normal to the 
rotating-trap plane; the energies are in units of $\hbar \omega$ and the 
lengths in units of the oscillator length $\Lambda=\sqrt{ \hbar/(M\omega) }$, with $M$ being the
fermion mass. The first and second terms express the LLL 
kinetic energy, $H_K$, and the third term represents the contact interaction, $H_{\rm int}$.

Our methodology integrating both numerical (e.g., fortran) and symbolic (algebraic, e.g.,
MATHEMATICA \cite{math}) languages consists of two steps: (1) numerical diagonalization of the 
Hamiltonian matrix problem employing the ARPACK solver \cite{arpack,arno51} of large-scale 
sparse eigenvalue problems, followed by step (2) where the numerically exact CI wave functions
\begin{align}
\Phi_{\rm CI} (z_1\sigma_1, \ldots , z_N\sigma_N) =
\sum_I c_{\rm CI}(I) \Psi_I(z_1\sigma_1, \ldots , z_N\sigma_N),
\label{phici}
\end{align}
are analyzed and processed using symbolic scripts targeting extraction of the corresponding exact
analytical wave functions. 

The basis Slater determinants that span the Hilbert space are 
\begin{align}
\Psi_I = {\rm Det}[ \varphi_{j_r}(z_s)\sigma_{j_r}(s) ]/ \sqrt{N!}, 
\label{detexd}
\end{align}
where $r,s=1,\ldots,N$, 
the LLL single-particle orbitals are
\begin{align}
\varphi_j(z)=z^{l_j} e^{-zz^*/2}/\sqrt{\pi l_j!},
\label{phi}
\end{align}
and $\sigma$ signifies an up ($\alpha$) or a down ($\beta$) spin. The master index $I$ 
counts the number of ordered arrangements (lists) $\{j_1,j_2,\ldots,j_N\}$  under the 
restriction that $1 \leq j_1 < j_2 <\ldots < j_N \leq K$; $K \in \mathbb{N}$ is chosen large 
enough to provide numerical convergence. Below, explicit mention of the Gaussian factor is omitted.

Step (2) starts with the rewriting of the CI wave function $\Phi_{\rm CI}$ in 
Eq.~(\ref{phici}) as 
\begin{align}
\Phi_{\rm alg} (z_1\sigma_1, \ldots , z_N\sigma_N) =
\sum_I c_{\rm alg}(I) \Psi_I(z_1\sigma_1, \ldots , z_N\sigma_N),
\label{phialg1}
\end{align}
where the replacement of the subscript ``CI'' by ``alg'' corresponds to the fact that, using the 
symbolic language code, one obtains an equivalent  multivariate homogeneous polynomial 
$\Phi_{\rm alg}$ with algebraic coefficients $c_{\rm alg}$; see the transcription of coefficients 
for $N=4$ and $N=9$ in Tables \ref{tn4l8} and \ref{tn940}
in the SM \cite{supp}.

Validation of our closed-form analytic wave functions (see below) is achieved via direct 
comparison of the numerical CI coefficients, $c_{\rm CI}$, with those in $\Phi_{\rm alg}$ 
[Eq.~(\ref{phialg1})], thus circumventing uncertainties, associated with the common use of wave 
function overlap \cite{laug83,simochap,jainbook,palm20,yoso98}, due to the van Vleck-Anderson
orthogonality catastrophe \cite{vlec36,ande67,kohn99,deml12,guzi14,ares18,gu19}.

Invariably, the symbolic code is able to simplify the derived multivariate polynomial 
in Eq.~(\ref{phialg1})
to the compact form of a product of a Vandermonde determinant (VDdet), $\prod_{i<j}^N (z_i-z_j)$, 
involving the space coordinates only, with a symmetric polynomial (under two-particle exchange) 
with mixed space and spin coordinates [see Eq.~(\ref{phialg}) below]. The factoring out of the 
VDdet reflects the fact that $\Phi_{\rm alg}$ represents a 0IE LLL state.

Using symbolic scripts, we verify further that the fully-algebraic $\Phi_{\rm alg}$
[Eq.~(\ref{phialg})] is indeed an eigenstate of the total spin, obeying the Fock condition 
\cite{fock40}. The final closed form expressions [see Eq.~(\ref{pol}) below] are derived for 
$N\leq 9$, but they are valid for any $N$, thus circumventing the CI numerical diagonalization 
of large matrices, which is not feasible for $N \geq \sim 10$.

For the CI diagonalization, a small perturbing term $V_P$ (e.g., a small trap deformation 
\cite{yann20}, or a small hard-wall boundary \cite{maca17}) needs to be added to the LLL 
Hamiltonian in Eq.~(\ref{H_lll}). This has a negligible influence on the numerical eigenvalues, 
but it is instrumental in lifting the degeneracies among the 0IE states, and thus produce CI 
states whose total spin $S$ is a good quantum number.

\noindent
\section{Targeted total spins and angular momenta}

For each size $N$, we provide analytic expressions for the maximum-spin ($S=S_z$) 0IE ground 
states with angular momenta $L=L_0+\Delta L$ [with $L_0=N(N-1)/2$] from $\Delta L=0$ (maximum
density droplet) to $\Delta L=N$ (first quasihole, 1QH); they form two families $A$ and $B$
(see Fig.~\ref{fspec} for an illustration).

Using $k$ to denote the number of spin-up fermions and $p$ that of spin-down fermions, and
focusing on the case with $k \geq p$ (or equivalently $p \leq N/2$), the states in both
families are associated with the same set of total spins specified as $S=S_z=(k-p)/2=N/2-p$.
Furthermore, given a pair $(k,p)$:

(A) The states in family $A$ have $\Delta L = p$, with
$\Delta L$ varying from 0 to $N/2$ for even $N$, and from 0 to $(N-1)/2$ for odd $N$.

(B) The states in family $B$ have $\Delta L = k$, with $\Delta L$ varying
from $N/2$ to $N$ for even $N$, and from $(N+1)/2$ to $N$ for odd $N$.

The states in family $A$ are unique ground states, whereas those in family $B$
are part of degenerate manifolds. (This degeneracy is lifted as described above.) 

\section{The exact 0IE LLL wave functions}

\subsubsection{Mathematical preliminary}

The~quantity $k$-subset($list$) is 
a subset containing exactly $k$ elements out of the set of $n$ elements (named $list$).
The number of $k$-subsets on $n$ elements is given by 
$ \left( \frac{n}{k} \right) =\frac{n!}{k!(n-k)!}$. The set represented by $list$ is 
taken to be a list of cardinally ordered positive integers. For example, there 
are 6 2-subsets when $list$=\{1,2,3,4\}, namely \{1,2\}, \{1,3\}, \{1,4\}, \{2,3\}, \{2,4\}, and 
\{3,4\}.

\subsubsection{General form of the 0IE LLL wave functions}

The compact algebraic expression has the general form 
\begin{align}
 \Phi_{\rm alg}\big( z_1 \chi(1),& \ldots,z_N \chi(N) \big)
 \propto \vdn (l_1,\ldots,l_N; z_1,\ldots,z_N) \nonumber \\
& \times  \sypa \big( z_1 \chi(1),\ldots,z_N \chi(N) \big),
\label{phialg}
\end{align}
where $\chi(i)$ denotes an up spin, $\alpha$, or down spin, $\beta$, and $i=1, \ldots, N$.

$\vdn$ is a Vandermonde determinant,
\begin{align}
\vdn ([l];[z]) = {\rm Det}[z_i^{l_j}] = \prod_{i<j}^N (z_i-z_j),
\label{vd} 
\end{align}
where $l_j=(j-1)$ and $i,j=1,2,\ldots,N$. The product of Jastrow factors above 
reflects the fact that the wave function in Eq.~(\ref{phialg}) is a 0IE eigenstate 
of the contact-interaction term, $H_{\rm int}$, in Eq.~(\ref{H_lll}).

Due to the fermionic symmetry of the $\Phi_{\rm alg}$, $\sypa$ has to be symmetric 
under the exchange of any pair of indices $i$ and $j$. Furthermore, $\sypa$ can be written as
\begin{align}
\sypa \big( z_1 \chi(1),\ldots,z_N \chi(N) \big)=
\sum_{m=1}^K \cp_m^o [z] \cz_m,
\label{pol} 
\end{align}
where $\cp_m^o$ (defined below) are homogeneous multivariate polynomials of order $o=p$
(family $A$) or $o=k$ (family $B$), and
\begin{align}
\cz_m = \alpha(i_1)\alpha(i_2)\ldots\alpha(i_k)\beta(j_{k+1})\ldots\beta(j_N),
\label{zeta}
\end{align}
is one of the $K=N!/(k!p!)$ distinct spin primitives having $k \leq N$ up and 
$p=N-k \leq k$ down spins. The set of indices $\{i_1,\ldots,i_k\}$ is the $m$th element 
($m=1,2,\ldots,K$) of the $k$-subsets of the cardinal list (top-level $list$, see below) 
specified as $list$=\{1,2,...,$N$\}. The set of indices 
$\{j_{k+1},\ldots,j_N\}$ is complementary to the $\{i_1,\ldots,i_k\}$ set.

The $\vdn$ [Eq.~(\ref{vd})] corresponds to a filling factor $\nu=1$, whereas the filling 
fraction corresponding to Eq.~(\ref{phialg}) [with $\sypa$ given in Eq.~(\ref{pol}) through 
polynomials of order $o$] is near $\nu=1$. These fractions are indeed the ones most likely to be 
accessed first in upcoming experiments \cite{palm20}.

\subsubsection{Algebraic expressions for the polynomials $\cp_m^o([z])$}

For each $S=S_z=(k-p)/2$, except when $k=p$ which has a single state, there exists a pair of
targeted LLL states, with one state of the pair belonging to family $A$ and the other to family $B$
(see Fig.~\ref{fspec} for an example).

\begin{figure*}[t]
\centering\includegraphics[width=15.5cm]{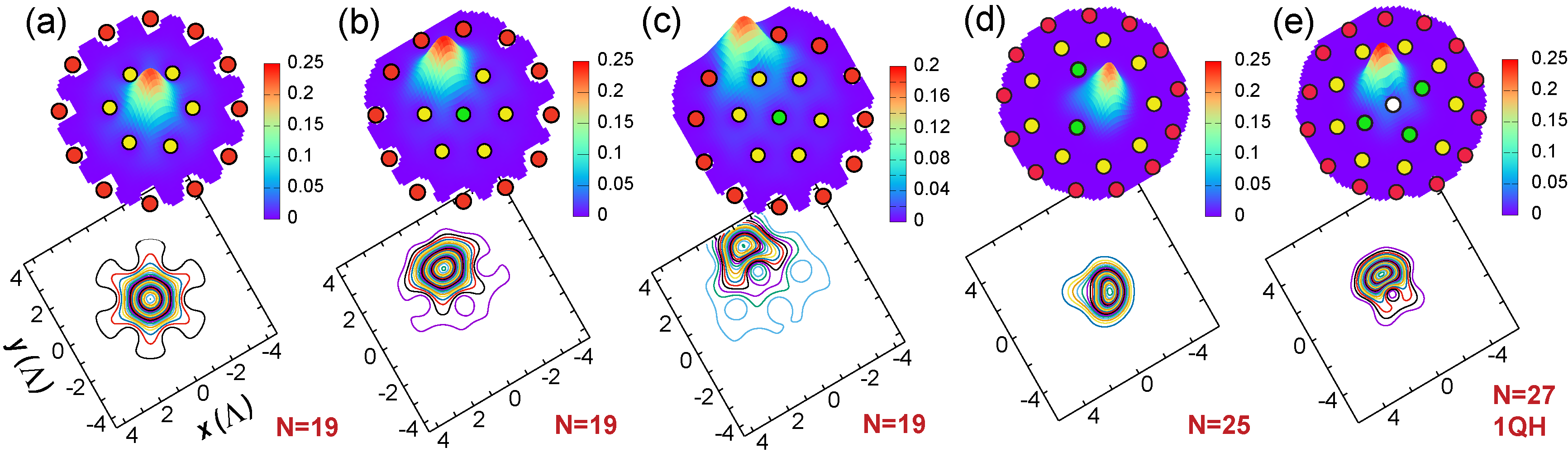}
\caption{
\textcolor{black}{
$N$th-order spin-unresolved correlations for $N$ LLL fermions. 
(a,b,c) $N=19$ with $L=180$, and $S=S_z=1/2$. (d) $N=25$ with $L=301$, and $S=S_z=23/2$. 
(e) $N=27$ with $L=378$, and $S=S_z=27/2$ (1QH state). Fixed fermions are marked by red dots for 
the outer ring, yellow dots for the middle ring, and green dots for the inner ring. The white dot 
for the 1QH state denotes the additional zero at the origin. Vertical axes: arbitrary units. 
See text for details. In the LLL, momentum correlations coincide with the spatial ones 
\cite{yann20}.}
}
\label{f982}
\end{figure*}

{\it Family A:\/} First, the following square matrices of rank $p$ (the number of 
spin-down fermions) need to be considered:
\begin{align}
{\rm M}_{q,m} = 
\left[
\begin{array}{ccc}
z_{i_1}-z_{j_{k+1}} & \dots & z_{i_1}-z_{j_N} \\
\vdots & \ddots & \vdots \\
z_{i_p}-z_{j_{k+1}} & \dots & z_{i_p}-z_{j_N} \\
\end{array}
\right],
\label{matqm}
\end{align}
where the dummy indices $i_1,\ldots,i_p$ 
\textcolor{black}{here} 
are associated with {\it spin-up\/} fermions, and
the set $\{i_1,\ldots,i_p\}$ denotes the $q$th subset among the $p$-subsets on a second-level 
$list$-2, with $list$-2 being the $m$th element among the $k$-subsets on the $\{1,2,\ldots,N\}$ 
top-level $list$. The number of $p$-subsets of any second-level $list$-2 is $K_2=k!/(p!(k-p)!)$, 
and thus the $q$ subscript runs from 1 to $K_2$. The set of indices $\{j_{k+1},\ldots,j_N\}$ is 
complementary to the $\{i_1,\ldots,i_k\}$ set, and thus it remains constant for a given $m$ index 
in the matrices defined in Eq.~(\ref{matqm}). (Recall that $k$ is the total number of spin-up 
fermions, and that $\{i_1,\ldots,i_k\}$ is also referred to as a second-level list.)

The expression for the polynomial is given by
\begin{align}
\cp_m^p([z])=\sum_{q=1}^{K_2} {\rm Perm}[{\rm M}_{q,m}],
\label{polm}
\end{align}
where the symbol ''Perm'' denotes a Permanent.

The analytic expressions of the states with $S_z < S$, in a given spin multiplicity $2S+1$, are 
obtained by repeated application of the spin lowering operator.

{\it Example\/.}  We consider the state associated with $N=5$, 
$S=S_z=1/2$, and $L=12$. Note that
$L_0=N(N-1)/2=10$ in the corresponding fully polarized case. 
There are $K=5!/(3!2!)=10$ spin primitives $\cz_m$, with $m=1,2,\ldots,10$; 
they correspond to the ten 3-subsets on the top-level $list$=$\{1,2,3,4,5\}$, i.e., 
$\{1,2,3\}$ ($m=1$), $\{1,2,4\}$ ($m=2$), $\{1,2,5\}$ ($m=3$), $\{1,3,4\}$ ($m=4$), 
$\{1,3,5\}$ ($m=5$), $\{1,4,5\}$ ($m=6$), $\{2,3,4\}$ ($m=7$), $\{2,3,5\}$ ($m=8$), 
$\{2,4,5\}$ ($m=9$), $\{3,4,5\}$ ($m=10$). 

Here $k=3$, $p=2$, and there are $K_2=3$ 2-subsets for each ($m$th) 3-subset listed above. 
$K_2=3$ is also the number of permanents entering in expression (\ref{polm}), i.e., $q=1,\dots,3$. 
Choosing $m=10$ as an example, the three 2-subsets are $\{3,4\}$, $\{3,5\}$, and 
$\{4,5\}$, and the three associated matrices ${\rm M}_{q,10}$ are given by:
\begin{align}
{\rm M}_{q,10} =
\left[
\begin{array}{ccc}
\eta(q,1)-z_1 & \eta(q,1)-z_2 \\
\eta(q,2)-z_1 & \eta(q,2)-z_2 \\
\end{array}
\right],
\label{matq10}
\end{align}
with with $q=1,2,3$; $\eta(1,1)=z_3$, $\eta(1,2)=z_4$, $\eta(2,1)=z_3$, $\eta(2,2)=z_5$, and
$\eta(3,1)=z_4$, $\eta(3,2)=z_5$.

An additional example is presented in the SM \cite{supp}.

{\it Family B:\/} Similarly, we found that the symmetric polynomials $\cp_m^k([z])$ related to 
the ground states of family $B$ consist always [for any $m$ in the summation of Eq.~(\ref{pol})] 
of a single permanent associated with a matrix of rank $k$ (the number of spin-up fermions). 
Namely 
\begin{align}
\cp_m^k([z])={\rm Perm}[{\rm M}_m^B],
\label{polbm}
\end{align}
with 
\begin{align}
& {\rm M}_m^B = \nonumber \\
& \left[
\begin{array}{cccccc}
z_{i_1}-z_{j_{k+1}} & \dots & z_{i_1}-z_{j_N} & z_{i_1}-z_{j_{N+1}} & \ldots & z_{i_1}-z_{j_{2k}} \\
\vdots & \ddots & \vdots & \vdots & \ddots & \vdots \\
z_{i_k}-z_{j_{k+1}} & \dots & z_{i_k}-z_{j_N} & z_{i_k}-z_{j_{N+1}} & \ldots & z_{i_k}-z_{j_{2k}} \\
\end{array}
\right].
\label{matbm}
\end{align}
Above, the set of indices $\{i_1,\ldots,i_k\}$ is the $m$th element of the 
$k$-subsets associated with the spin-up fermions [see Eq.~(\ref{zeta})]. Because $k>p$, the 
complimentary set of the $p$ spin-down indices $\{j_{k+1},\ldots,j_N\}$ has been expanded to 
contain exactly $k$ elements, through the introduction of virtual fermion coordinates such that 
$z_{j_s}=0$ for all $s>N$; see specific matrices ${\rm M}_m^B$, as well as a comparison with 
the wave functions in Ref.~\cite{brey96}, in the Appendix and the SM \cite{supp}. 

Note that the first quasi-hole state (1QH) \cite{laug83,maca17} coincides with the analytic
expression associated with family $B$ above for $L=L_0+N$.

\section{Higher-order correlations}

We used the analytic wave functions above to calculate spin-unresolved higher-oder correlations 
for $N=19$, 25, and 27 fermions; see Fig.~\ref{f982} (for completeness, see Fig.\  
\ref{figcomp} for $N=9$ in the SM \cite{supp}).
The $n$-body correlations for spinful fermions were defined in detail in Sec. II C of 
Ref.~\cite{yann20}. For the $N$-body $\Phi_{\rm alg}$ [Eq.~(\ref{phialg})], 
the spatial $n$-body correlation is given in a compact form by
\begin{align}
 \cg^n & (N)= (1-\delta_{n,N}) 
\int | \Phi_{\rm alg} \big( z_1 \chi(1), \ldots, z_N \chi(N) \big)|^2 \times \nonumber \\ 
& dz_{n+1} d\chi(n+1) \ldots dz_N d\chi(N) 
+ \delta_{n,N} |\Phi_{\rm alg}|^2,
\label{nbody}
\end{align}
\textcolor{black}{
with $n=2, \dots, N$. $\cg^n (N)$ gives the conditional probablility to find particles 
$n,\ldots,N$ anywhere, for prespecified (fixed) locations of particles $1,\ldots,n-1$ with 
predetermined (resolved) or unspecified (unresolved) spins.  
}

\textcolor{black}{
For $N=19$, Figs.~\ref{f982}(a,b,c) display structured $N$th-order correlations for the 
spin state with $S=S_z=1/2$ and total angular momentum $L=180$. Extending 
Ref.~\cite{yann20}, we found similar crystalline structures 
also in the $N$th-order correlations of the associated 
fully polarized, single VDdet state with $S=19/2$, $S_z=19/2$, and $L_0=171$ 
(Pauli-exclusion-only case, experimentally investigated \cite{holt20}). Fig.~\ref{f982}(d) 
displays the structured $N$th-order correlation for $N=25$ with $L=301$ and $S=S_z=23/2$,
whereas Fig.~\ref{f982}(e) presents the structured $N$th-order correlations for the 1QH for
$N=27$ (with $L=378$ and $S=S_z=27/2$). The implied intrinsic geometric structure (UCWM)
in Fig.~\ref{f982} is a polygonal triple ring $(n_1,n_2,n_3)$ 
of localized fermions (with $n_1+n_2+n_3=N$); specifically (1,6,12), (3,9,13), and (4,9,14)
for Figs.~\ref{f982}(a,b,c), Fig.~\ref{f982}(d), and Fig.~\ref{f982}(e), respectively. 
We note that in the LLL neighbohood of $\nu=1$ (expected in experiments with 
trapped ultracold fermions \cite{palm20}), the intrinsic ring geometry can be probed only with 
higher-order correlations. Indeed in this case, the second-order correlations are structureless;
see the findings for $N=4$ [(0,4) single ring] and $N=6$ [(1,5) double ring] in 
Ref.~\cite{yann20}. 
}

\section{Conclusion}

A novel approach for deriving exact closed-form analytic expressions for the wave functions 
(beyond the Jastrow-factors paradigm) of an assembly of 2D contact-interacting spinful LLL 
fermions (for any $N$) was introduced and validated. Such expressions require as input only the 
three parameters $N$ (number of particles), $L$ (total angular momentum), and $S$ (total spin).  
Examples were presented for two families of zero-interaction-energy 
states, from the maximum density droplet to the first quasihole in the neighborhood of $\nu=1$. 
Ensuing theoretical predictions for higher-order momentum correlations for $N=19$, 25, and 27,
revealing intrinsic polygonal, multi-ring crystalline configurations, could be tested with 
ultracold-atom experiments in rotating traps  simulating spinful quantum Hall physics, 
including LLL skyrmions. The present approach can be extended to the neighborhood of any 
$\nu=1/m$ that starts with a Laughlin wave function.

\section{Acknowledgments}
 
This work has been supported by a grant from the Air Force Office of Scientific Research (AFSOR) 
under Award No. FA9550-21-1-0198. Calculations were carried out at the GATECH Center for 
Computational Materials Science.

\appendix*

\section{Comparison with the symmetric polynomials for quantum skyrmions in Ref.\ \cite{brey96}}

We compare here with the symmetric polynomials for the seed skyrmions specified in Eq.~(6)
of Ref.\ \cite{brey96} or Eq.~(2) in Ref.\ \cite{jain96}.

Omitting the trivial Gaussian functions, these polynomials are given by the single formula
\begin{align}
\Phi^{\rm sk,MFB}_p = \sum_{m=1}^K z_{i_1}z_{i_2}...z_{i_k} \cz_m,
\label{phimfb}
\end{align}  
where $\cz_m$ are the spin primitives defined in Eq.\ (\ref{zeta}), and the superscript MFB stands
for MacDonald-Fertig-Brey. The index $m$ runs also over the $k$-subsets $\{i_1,\ldots,i_k\}$ 
of the $list=\{1,2,3,\ldots,N\}$, $k=N\!\uparrow$ being the number of spin-up fermions, with 
$p=N-k=N\!\downarrow$ being that of the spin-down fermions. The number of $k$-subsets is
$K=N!/(k!p!)$. As is the case in Ref.\ \cite{brey96}, one can take the index $m$ as running over 
the $p$-subsets associated with the spin-down fermions, because there is a one-to-one 
correspondence to the $k$-subsets of the spin-up fermions. Note that Ref.\ \cite{brey96} 
(Ref.\ \cite{jain96}) uses the capital letter $K$ ($R$) in place of our $p$.

We consider the case $N=5$, $k=4$, $p=1$, $S=S_z=3/2$, and $\Delta L = 4$, belonging to family $B$
in our exposition.

According to Eq.~(\ref{phimfb}), the corresponding MFB symmetric polynomial is
\begin{align}
\begin{split}
\Phi^{\rm sk,MFB}_{p=1}= & z_1 z_2 z_3 z_4 \cz_1  + z_1 z_2 z_3 z_5 \cz_2 + 
z_1 z_2 z_4 z_5 \cz_3 +\\
& z_1 z_3 z_4 z_5 \cz_4  + z_2 z_3 z_4 z_5 \cz_5.
\end{split}
\label{phimfb5}
\end{align} 

The corresponding exact symmetric polynomial derived in this paper is given by Eqs.\ (\ref{pol}) 
and (\ref{polbm}), namely
\begin{align}
\Phi^{\rm exact}_{\rm sym}(N=5, N\!\uparrow=4, \Delta L=4) = 
\sum_{m=1}^5 {\cal P}^4_m[z] {\cal Z}_m,
\label{phin5nup4}
\end{align}

Expanding the permanents, one obtains for the space-only polynomials ${\cal P}^4_m[z]$ 
above (with $m=1,\ldots,5$, in front of the $\cz_m$ spin primitives): 
\begin{align}
\begin{split}
 {\cal P}^4_m[z]= & c_1 z_1 z_2 z_3 z_4 + c_2 z_1 z_2 z_3 z_5 + c_3 z_1 z_2 z_4 z_5 + \\
& c_4 z_1 z_3 z_4 z_5 + c_5 z_2 z_3 z_4 z_5,
\end{split}
\label{p4m}
\end{align}
with $c_i=4$ when $i=m$ and $c_i=-1$ otherwise. 

The polynomial in Eq.\ (\ref{phin5nup4}) is clearly different from the MFB one 
[Eq.\ (\ref{phimfb5})]. We verified that the wave functions derived here are eigenfunctions of 
the square, $\hat{S}^2$, of the total-spin operator [with eigenvalue 15/4 and $S=3/2$ for the 
case in this Appendix], whereas the MFB ones are not (see also Ref.\ \cite{abol97}); for details 
see Ref.\ \cite{supp}.

\newpage

\setcounter{table}{0}
\makeatletter 
\renewcommand{\thefigure}{SF\@arabic\c@figure}
\renewcommand{\thetable}{ST\@Roman\c@table}
\renewcommand{\theequation}{S\@arabic\c@equation}
\makeatother

\begin{widetext}

\vspace{2cm}
\begin{center}
{\bf{\Large SUPPLEMENTAL MATERIAL \\
~~~~\\
Exact closed-form analytic wave functions in two dimensions: Contact-interacting fermionic
spinful ultracold atoms in a rapidly rotating trap}}\\
~~~~~\\
{\Large Constantine Yannouleas$^\dagger$ and Uzi Landman$^*$\\
{\it School of Physics, Georgia Institute of Technology,
             Atlanta, Georgia 30332-0430}}
\end{center}
~~~~~~\\
~~~~~~\\
~~~~~~\\
~~~~~~\\
~~~~~~\\
~~~~~~\\
\noindent 
{\bf Contents:\\
~~~~~\\
1) Contrast with the symmetric polynomials for skyrmions in earlier literature, p.~9\\
~~~~~\\
2) Additional specific examples of the compact analytic wave functions derived in the
main text, p.~10\\
~~~~\\ 
3) Tables illustrating, for $N=4$ and $N=9$, the correspondence between numerical CI 
and algebraic coefficients in the determinantal expansions of $\Phi_{\rm CI}$ and 
$\Phi_{\rm alg}$, p.~11 \\
~~~~~~~\\
4) Figure displaying additional higher-order correlations for $N=9$ LLL fermions, p.12
~~~~~~~\\
~~~~~~~\\
~~~~~~~\\
~~~~~~\\
~~~~~~\\
}
$^\dagger$Constantine.Yannouleas@physics.gatech.edu\\
$^*$Uzi.Landman@physics.gatech.edu
\newpage

\noindent
\noindent
\begin{center}
{\LARGE {\bf Comparison with symmetric polynomials for skyrmions in previous literature}}
\end{center}
\label{comp}

We compare here with the symmetric polynomials for the seed skyrmions specified in Eq.~(6)
of (R1) A. H. MacDonald, H. A. Fertig, and Luis Brey, Skyrmions without Sigma Models in Quantum 
Hall Ferromagnets, Phys. Rev. Lett. {\bf 76}, 2153 (1996) 
(\url{https://doi.org/10.1103/PhysRevLett.76.2153});
see also Eq.~(2) in (R2) R.K. Kamilla, X.G. Wu, and J.K. Jain, Skyrmions in the fractional quantum 
Hall effect, Solid State Commun. {\bf 99}, 289 (1996) 
(\url{https://doi.org/10.1016/0038-1098(96)00126-3}).

Omitting the trivial Gaussian functions, these polynomials are given by the single formula
\begin{align}
\Phi^{\rm sk,MFB}_p = \sum_{m=1}^K z_{i_1}z_{i_2}...z_{i_k} \cz_m,
\label{phimfb2}
\end{align}  
where $\cz_m$ are the spin primitives defined in Eq.~(9) of the main text, i.e., 
\begin{align}
\cz_m = \alpha(i_1)\alpha(i_2)\ldots\alpha(i_k)\beta(j_{k+1})\ldots\beta(j_N).
\label{cz}
\end{align} 
The index $m$ runs also over the $k$-subsets $\{i_1,\ldots,i_k\}$ 
of the $list=\{1,2,3,\ldots,N\}$, $k=N\!\uparrow$ being the number of spin-up fermions, with 
$p=N-k=N\!\downarrow$ being that of the spin-down fermions. The number of $k$-subsets is
$K=N!/(k!p!)$. As is the case in Ref.\ R1, one can take the index $m$ as running over the 
$p$-subsets associated with the spin-down fermions, because there is a one-to-one correspondence 
to the $k$-subsets of the spin-up fermions. Note that Ref.\ R1 (Ref.\ R2) uses the capital letter 
$K$ ($R$) in place of our $p$.

We present here a comparison for the case $N=5$, $k=4$, $p=1$, $S=S_z=3/2$, and $\Delta L = 4$, a 
state belonging to family B in our exposition.

According to Eq.~(\ref{phimfb2}), the corresponding MacDonald-Fertig-Brey (MFB) symmetric
polynomial is
\begin{align}
\Phi^{\rm sk,MFB}_{p=1} = 
z_1 z_2 z_3 z_4 \cz_1  + z_1 z_2 z_3 z_5 \cz_2 + z_1 z_2 z_4 z_5 \cz_3 +
z_1 z_3 z_4 z_5 \cz_4  + z_2 z_3 z_4 z_5 \cz_5.
\label{phimfb52}
\end{align} 

The corresponding polynomial derived in this paper is given by Eqs.~(8) and (13) in the main 
text. Expanding the permanent, one obtains for the polynomial with $k=4$ and $m=1$ (in front of
the $\cz_1$ spin primitive):
\begin{align}
{\cal P}^4_1[z]=4 z_1 z_2 z_3 z_4 - z_1 z_2 z_3 z_5 - z_1 z_2 z_4 z_5 - 
 z_1 z_3 z_4 z_5 - z_2 z_3 z_4 z_5.
\label{p41}
\end{align} 

For the ${\cal P}^4_2$ polynomial in front of $\cz_2$, one obtains similarly:
\begin{align}
{\cal P}^4_2[z]=-z_1 z_2 z_3 z_4 + 4 z_1 z_2 z_3 z_5 - z_1 z_2 z_4 z_5 - 
 z_1 z_3 z_4 z_5 - z_2 z_3 z_4 z_5.
\label{p42}
\end{align}

In general, one has 
\begin{align}
\Phi^{\rm exact}_{\rm sym}(N=5, N\!\uparrow=4, \Delta L=4) = 
\sum_{m=1}^5 {\cal P}^4_m[z] {\cal Z}_m,
\label{phin5nup42}
\end{align}
with 
\begin{align}
{\cal P}^4_m[z]=c_1 z_1 z_2 z_3 z_4 + c_2 z_1 z_2 z_3 z_5 + c_3 z_1 z_2 z_4 z_5 + 
c_4 z_1 z_3 z_4 z_5 + c_5 z_2 z_3 z_4 z_5,
\label{p4m2}
\end{align}
and $c_i=4$ when $i=m$ and $c_i=-1$ otherwise.

We note that the expressions associated with the $\cz_i$, $i=1,\ldots,5$ in the MFB polynomial 
consist only of a single term with a numerical factor +1 in front. This contrasts with our 
expressions in Eqs.~(\ref{p41}), (\ref{p42}), and (\ref{p4m2}) which have five terms each with 
factors of +4 and -1 in front of them.

For $p$ (spin-down fermions) $>$ $k$ (spin-up fermions), the MFB expression in Eq.~(\ref{phimfb2})
is associated with a negative total-spin projection $S_z=(k-p)/2<0$. In this case, the indices for
the corresponding wave function in this paper are found by reversing all $N$ spins, i.e., by 
considering the case with $p \rightarrow k$, $k \rightarrow p$, and $S_z=|(k-p)/2|$.  

Using our algebraic scripts, we readily verified that the wave functions derived in this work are 
indeed eigenfunctions of the total-spin square operator [with eigenvalue 15/4 and $S=3/2$ for the 
case in this section], whereas the MFB ones are not (as indeed has been discussed by M. Abolfath 
{\it et al.\/}, Phys. Rev. B {\bf 56}, 6795 (1997) 
(\url{https://doi.org/10.1103/PhysRevB.56.6795}).

In particular, applying the spin-square, $\hat{S}^2$, operator, one gets
\begin{align}
\big( \hat{S}^2-\frac{15}{4} \big) \Phi^{\rm exact}_{\rm sym}(N=5, N\!\uparrow=4, \Delta L=4)
= 0.
\label{s2exact}
\end{align} 

On the contrary, for the MFB wave function, one gets
\begin{align}
\big( \hat{S}^2-\frac{15}{4} \big) \Phi^{\rm sk,MFB}_{p=1} = 
 (z_1 z_2 z_3 z_4 + z_1 z_2 z_3 z_5 + z_1 z_2 z_4 z_5 + z_1 z_3 z_4 z_5 + z_2 z_3 z_4 z_5)
\sum_{m=1}^5 {\cal Z}_m.
\label{s2mfb}
\end{align}

\noindent
=====================================================

\begin{center}
{\LARGE {\bf Additional examples for the wave functions in families $A$ and $B$}}
\end{center}
\label{exam}

{\bf Family $A$.}
As another example from family $A$, we consider the spin singlet state associated with $N=4$,
$S=0$, $S_z=0$, and $L=8$. Note that $L_0=N(N-1)/2=6$ in the corresponding fully spin-polarized
case. There are $K=4!/(2!2!)=6$ spin primitives $\cz_m$, with $m=1,2,\ldots,6$; they correspond to
the six 2-subsets on the top-level $list=\{1,2,3,4\}$, i.e.,
$\{1,2\}$ ($m=1$), $\{1,3\}$ ($m=2$), $\{1,4\}$ ($m=3$), $\{2,3\}$ ($m=4$),
$\{2,4\}$ ($m=5$), $\{3,4\}$ ($m=6$).

Because $k=2$ and $p=2$, there is only one ($K_2=1$) 2-subset for each ($m$th) 2-subset listed
above. $K_2=1$ is also the number of permanents entering in expression (11) of the main text, 
i.e., the index $q$ takes only the value of one. Choosing $m=2$ as an example, the single matrix
${\rm M}_{1,2}$ is:
\begin{align}
{\rm M}_{1,2} =
\left[
\begin{array}{ccc}
z_1-z_2 &  z_1-z_4 \\
z_3-z_2 &  z_3-z_4 \\
\end{array}
\right].
\label{mat110}
\end{align}

{\bf Family $B$. First example.\/}
As a first example from family $B$, we consider the spin $S=S_z=3/2$ state associated with 
$N=9$ and $L=42$. Note that $L_0=N(N-1)/2=36$. There are $k=6$ up spins, $p=3$ down spins, 
and $K=9!/(6!3!)=84$ spin primitives $\cz_m$, with $m=1,2,\ldots,84$.

Focusing on the $m=1$ spin primitive, we have a subset of indices $\{i_1,i_2,\ldots,i_6\}$ 
for the spin up fermions and a subset of indices $\{j_7,j_8,j_9\}$ for the spin down fermions.
The subset of the spin down indices needs to be augmented by introducing three additional 
indices $\{j_{10},j_{11},j_{12}\}$, which specify virtual fermions with 
$z_{j_{10}}=z_{j_{11}}=z_{j_{12}}=0$.
Then the corresponding matrix ${\rm M}_1^B$ [see Eq.~(14) in the main text] is given by
\begin{align}
{\rm M}_1^B =
\left[
\begin{array}{cccccc}
z_{i_1}-z_{j_7} & z_{i_1}-z_{j_8} & z_{i_1}-z_{j_9} & z_{i_1} & z_{i_1} & z_{i_1} \\
z_{i_2}-z_{j_7} & z_{i_2}-z_{j_8} & z_{i_2}-z_{j_9} & z_{i_2} & z_{i_2} & z_{i_2} \\
z_{i_3}-z_{j_7} & z_{i_3}-z_{j_8} & z_{i_3}-z_{j_9} & z_{i_3} & z_{i_3} & z_{i_3} \\
z_{i_4}-z_{j_7} & z_{i_4}-z_{j_8} & z_{i_4}-z_{j_9} & z_{i_4} & z_{i_4} & z_{i_4} \\
z_{i_5}-z_{j_7} & z_{i_5}-z_{j_8} & z_{i_5}-z_{j_9} & z_{i_5} & z_{i_5} & z_{i_5} \\
z_{i_6}-z_{j_7} & z_{i_6}-z_{j_8} & z_{i_6}-z_{j_9} & z_{i_6} & z_{i_6} & z_{i_6} \\
\end{array}
\right].
\label{matbm2}
\end{align}

{\bf Family $B$. Second example.\/}
As a second example from family $B$, we consider the 1QH state for $N=9$. In this case,
$S=S_z=9/2$ and $L=45$. Note that $L_0=N(N-1)/2=36$. There are $k=9$ up spins, $p=0$ down spins,
and $K=9!/(9!0!)=1$ spin primitive $\cz_1$.

For this single spin primitive, we have a subset of indices $\{i_1,i_2,\ldots,i_9\}$
for the spin up fermions and an empty subset of indices for the spin down fermions.
The subset of the spin down indices needs to be augmented by introducing nine additional 
indices $\{j_1,j_2,\ldots,j_9\}$, which specify virtual fermions with 
$z_{j_1}=z_{j_2}=\ldots =z_{j_9}=0$. Then the corresponding matrix ${\rm M}_1^B$ [see Eq.~(14) 
in the main text] is given by
\begin{align}
{\rm M}_1^B =
\left[
\begin{array}{ccccccccc}
z_{i_1} & z_{i_1} & z_{i_1} & z_{i_1} & z_{i_1} & z_{i_1} & z_{i_1} & z_{i_1} & z_{i_1} \\
z_{i_2} & z_{i_2} & z_{i_2} & z_{i_2} & z_{i_2} & z_{i_2} & z_{i_2} & z_{i_2} & z_{i_2} \\
z_{i_3} & z_{i_3} & z_{i_3} & z_{i_3} & z_{i_3} & z_{i_3} & z_{i_3} & z_{i_3} & z_{i_3} \\
z_{i_4} & z_{i_4} & z_{i_4} & z_{i_4} & z_{i_4} & z_{i_4} & z_{i_4} & z_{i_4} & z_{i_4} \\
z_{i_5} & z_{i_5} & z_{i_5} & z_{i_5} & z_{i_5} & z_{i_5} & z_{i_5} & z_{i_5} & z_{i_5} \\
z_{i_6} & z_{i_6} & z_{i_6} & z_{i_6} & z_{i_6} & z_{i_6} & z_{i_6} & z_{i_6} & z_{i_6} \\
z_{i_7} & z_{i_7} & z_{i_7} & z_{i_7} & z_{i_7} & z_{i_7} & z_{i_7} & z_{i_7} & z_{i_7} \\
z_{i_8} & z_{i_8} & z_{i_8} & z_{i_8} & z_{i_8} & z_{i_8} & z_{i_8} & z_{i_8} & z_{i_8} \\
z_{i_9} & z_{i_9} & z_{i_9} & z_{i_9} & z_{i_9} & z_{i_9} & z_{i_9} & z_{i_9} & z_{i_9} \\
\end{array}
\right].
\label{matbm3}
\end{align}

When expanded, the associated permanent yields one term only, namely,
$9! z_{i_1}z_{i_2}z_{i_3}z_{i_4}z_{i_5}z_{i_6}z_{i_7}z_{i_8}z_{i_9}$,
and thus the 1QH state here agrees with the analytic form introduced by Laughlin;
see \url{https://doi.org/10.1103/PhysRevLett.50.1395}.\\

\noindent
==================================================

\begin{table*}[t]
\caption{\label{tn4l8}
The 15 dominant numerical CI coefficients, $c_{\rm CI}(J)$, in the CI expansion of the relative 
LLL ground state, and the corresponding extracted algebraic ones, $c_{\rm alg}(J)$, 
for $N=4$ fermions with total angular momentum $L=8$ (first 0IE state) in the ($S=S_z=0$) 
spin sector of singlet states. The spinful-fermions Slater determinants are specified through the 
set of single-particle angular momenta and spins, 
$(l_1\uparrow,l_2\uparrow,l_3\downarrow,l_4\downarrow)$. The converged CI expansion included a 
much larger set of basis Slater determinants, but naturally the dominant ones are only relevant, 
namely those with coefficients $|c_{\rm CI}(I)|> 0.001$.}  
\begin{ruledtabular}
\begin{tabular}{r|c|r|c|c}
$J$ & $c_{\rm CI} (J)$ & $c_{\rm alg} (J)$ & 
$(l_1\uparrow,l_2\uparrow,l_3\downarrow,l_4\downarrow)$ &
$\sum_{i=1}^4 l_i $ \\ \hline
1 & -0.28005598 &  $ \frac{2}{ \sqrt{51} } $ & (0,1,3,4)  & 8 \\
2 &  0.34299719 & $ -\sqrt{ \frac{2}{17} } $ & (0,2,2,4)  & 8 \\
3 & -0.14002799 & $  \frac{1}{\sqrt{51}  } $ & (0,3,1,4)  & 8 \\
4 & -0.29704427 & $  \sqrt{ \frac{3}{34} } $ & (0,3,2,3)  & 8 \\
5 &  0.14002799 & $ -\frac{1}{\sqrt{51}  } $ & (0,4,1,3)  & 8 \\
6 & -0.24253563 & $  \frac{1}{\sqrt{17}  } $ & (1,2,1,4)  & 8 \\
7 & -0.17149861 & $  \frac{1}{\sqrt{34}  } $ & (1,2,2,3)  & 8 \\
8 &  0.14002799 & $ -\frac{1}{\sqrt{51}  } $ & (1,3,0,4)  & 8 \\
9 &  0.42008405 & $ -\sqrt{ \frac{3}{17} } $ & (1,3,1,3)  & 8 \\
10 & -0.14002799 & $  \frac{1}{\sqrt{51} } $ & (1,4,0,3)  & 8 \\
11 & -0.24253562 & $  \frac{1}{\sqrt{17} } $ & (1,4,1,2)  & 8 \\
12 & -0.29704427 & $  \sqrt{ \frac{3}{34} } $ & (2,3,0,3)  & 8 \\
13 & -0.17149861 & $  \frac{1}{\sqrt{34}  } $ & (2,3,1,2)  & 8 \\
14 &  0.34299718 & $ -\sqrt{ \frac{2}{17} } $ & (2,4,0,2)  & 8 \\
15 & -0.28005598 & $  \frac{2}{\sqrt{51}  } $ & (3,4,0,1)  & 8 \\
\end{tabular}
\end{ruledtabular}
\end{table*}

\begin{table*}[t]
\caption{\label{tn940}
A sample of the 1551 dominant numerical CI coefficients, $c_{\rm CI}(J)$, in the CI expansion of 
the relative LLL ground state, and the corresponding extracted algebraic ones, $c_{\rm alg}(J)$, 
for $N=9$ fermions with total angular momentum $L=40$ (first 0IE state) in the ($S=S_z=1/2$) 
spin sector. The spinful-fermions Slater determinants are specified through the 
set of single-particle angular momenta and spins, 
$(l_1\uparrow,l_2\uparrow,l_3\uparrow,l_4\uparrow,l_5\uparrow,
l_6\downarrow,l_7\downarrow,l_8\downarrow,l_9\downarrow)$.
The converged CI expansion included a much larger set of basis Slater determinants, but naturally 
the dominant ones are only relevant, namely those with coefficients $|c_{\rm CI}(I)|> 0.001$.}  
\begin{ruledtabular}
\begin{tabular}{r|c|r|c|c}
$J$ & $c_{\rm CI} (J)$ & $c_{\rm alg} (J)$ & 
$(l_1\uparrow,l_2\uparrow,l_3\uparrow,l_4\uparrow,l_5\uparrow,
l_6\downarrow,l_7\downarrow,l_8\downarrow,l_9\downarrow)$ &
$\sum_{i=1}^9 l_i $ \\ \hline
1 & -0.10319005  & $ 4 \sqrt{ \frac{5}{7513} }  $ & (0,1,2,3,4,6,7,8,9)  & 40 \\
2 &  0.11303906  & $ -4\sqrt{ \frac{6}{7513} }  $ & (0,1,2,3,5,5,7,8,9)  & 40 \\
3 & -0.02063800  & $  \frac{4}{\sqrt{37565}  }  $ & (0,1,2,3,6,4,7,8,9)  & 40 \\
4 & -0.10465386  & $  \frac{24}{\sqrt{52591} }  $ & (0,1,2,3,6,5,6,8,9)  & 40 \\
5 &  0.02063800  & $ -\frac{4}{\sqrt{37565}  }  $ & (0,1,2,3,7,4,6,8,9)  & 40 \\
6 &  0.09789477  & $ -6\sqrt{ \frac{2}{7513} }  $ & (0,1,2,3,7,5,6,7,9)  & 40 \\
$\vdots$ & $\vdots$ & $\vdots$ & $\vdots$ & $\vdots$ \\
1546 & 0.00872117  & $  -\frac{2}{ \sqrt{52591} }  $ & (4,5,6,7,9,0,1,3,5) & 40 \\
1547 & 0.00827362  & $ -3\sqrt{ \frac{2}{262955} } $ & (4,5,6,7,9,0,2,3,4) & 40 \\
1548 & -0.02136239 & $ 2\sqrt{ \frac{6}{52591} }   $ & (4,5,6,8,9,0,1,2,5) & 40 \\
1549 & -0.01103148 & $ 4\sqrt{ \frac{2}{262955} }  $ & (4,5,6,8,9,0,1,3,4) & 40 \\
1550 & 0.02527631  & $ -2\sqrt{ \frac{6}{37565} }  $ & (4,5,7,8,9,0,1,2,4) & 40 \\
1551 & -0.02063801 & $ \frac{4}{ \sqrt{37565} }    $ & (4,6,7,8,9,0,1,2,3) & 40 \\
\end{tabular}
\end{ruledtabular}
\end{table*}


\begin{figure*}[t]
\centering\includegraphics[width=15.8cm]{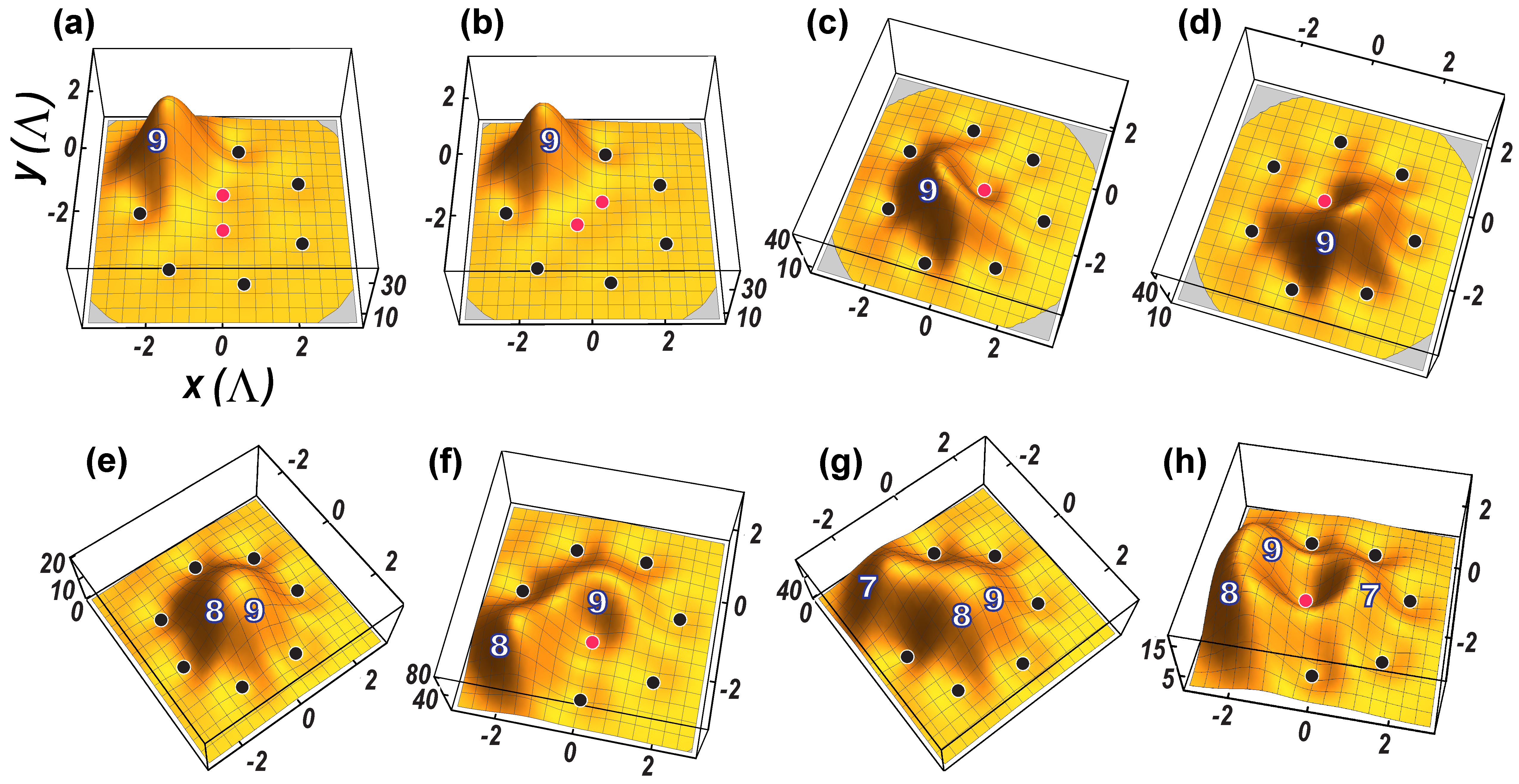}
\caption{
Higher-order correlations for $N=9$ ultracold LLL fermions. (a-d) 9th-order correlations for
the 0IE lowest-spin state ($S=1/2$, $S_z=1/2$) with total angular momentum
$L=40$. Similar 9th-order correlations are found for the fully polarized state 
($S=S_z=9/2$) with $L_0=36$.
(e,f) 8th-order correlations for the fully polarized state ($S=9/2$, $S_z=9/2$) with $L_0=36$,
whose space part is a pure Vandermonde determinant.
(g,h) 7th-order correlations for the fully polarized state ($S=9/2$, $S_z=9/2$) with $L_0=36$.
The implied intrinsic geometric structure (ultracold Wigner molecule) is a (2,7) double ring, 
with 2 fermions in the inner ring and 7 fermions in the outer ring. The fixed fermions are marked 
by solid dots (black for those in the outer ring and red for those
in the inner ring). The white numbers denote the remaining, beyond the fixed ones, fermions.
Vertical axes: arbitrary units.
}
\label{figcomp}
\end{figure*}

\end{widetext}

\end{document}